\documentclass{elsart5p}

\usepackage{natbib}

\usepackage{graphicx}

\usepackage{amssymb}

\begin{document}

\begin{frontmatter}



\title{Pulse variation of the optical emission of Crab pulsar}


\author[sao]{Karpov, S.}
\author[sao]{Beskin, G.}
\author[sai]{Biryukov, A.}
\author[sao]{Plokhotnichenko, V.}
\author[sao]{Debur, V.}
\author[nui]{Shearer, A.}

\address[sao]{Special Astrophysical Observatory of Russian Academy of Sciences,
Russia}
\address[sai]{Sternberg Astronomical Institute of Moscow State University, Moscow, Russia}
\address[nui]{National University of Ireland, Galway, Ireland}

\begin{abstract}
  The stability of the optical pulse of the Crab pulsar is analyzed based on
  the 1 $\mu$s resolution observations with the Russian 6-meter and William
  Hershel telescopes equipped with different photon-counting detectors. The
  search for the variations of the pulse shape along with its arrival time
  stability is performed. Upper limits on the possible short time scale free
  precession of the pulsar are placed. The evidence of pulse time of arrival
  (TOA) variations on 1.5-2 hours time scale is presented, along with evidence
  of small light curve (shape and separation of main and secondary peaks)
  changes between data sets, on time scale of years. Also, the fine structure
  of the main pulse is studied.
\end{abstract}

\begin{keyword}
Pulsars \sep Photometric, polarimetric, and spectroscopic instrumentation
\PACS 97.60.Gb \sep 95.55.Qf
\end{keyword}

\end{frontmatter}


\section{Introduction}

Over the last 30 years the Crab pulsar has been extensively studied.  The
reasons for it are clear -- it is the brightest pulsar seen in optics, it is
nearby and young.  However, the most popular groups of contemporary theories of
the Crab high-energy emission, the ``polar cap'' \citep{daugherty} and ``outer
gap'' \citep{cheng} ones, can't explain the whole set of observational data.

One of the main properties of the Crab emission is the relatively high
stability of its optical pulse shape despite the secular decrease of the
luminosity, related to the spin rate decrease \citep{pacini,nasuti}.

At the same time the pulsars in general and the Crab itself are unstable. The
instabilities manifest themselves as glitches, likely related to the changes of
the neutron star crust, timing noise, powered by the collective processes in
the superfluid internal parts of it, magnetospheric instabilities, results of
the wisps around the pulsar, precession, etc. All these factors may influence
the optical pulse structure and change it on various time scales, both in
periodic and stochastic way.

However, it has been found early that the variations of the Crab optical light
curve, in contrast with the radio ones, are governed by the Poissonian
statistics \citep{Kristian}.  A number of observations show the absence of
non-stationary effects in the structure, intensity and the duration of the Crab
optical pulses, and the restrictions on the regular and stochastic fine
structure of its pulse on the time scales from 3$\mu$s to 500$\mu$s
\citep{beskin,percival_1993}, the fluctuations of the pulse intensity
\citep{Kristian}.

Along with the increase of the observational time spans and the accuracy of
measurements, small changes of the optical pulse intensity, synchronous with
the giant radio pulses, have been detected \citep{shearer}. Also, the evidence
for the short time scale precession of the pulsar has been found by studying
its optical light curve \citep{cadez_2001}.

All this raises the importance of monitoring the Crab optical emission with
high time resolution.

The article is organized as follows. In Section 2 we briefly describe the
observation process and instruments used, in Section 3 the method of phase
stability analysis is described and used to study the Dec 1999 and Jan 2007
data sets, in Section 4 the light curves of different data sets are compared,
in Section 5 the possible fine structure of the main pulse peak is discussed,
and  Section 6 gives the conclusions.

\section{Observations and data reduction}

\begin{table*}[t]
\caption{Log of observations}
\label{table_observations}
\begin{tabular}{ccccc}
\hline\noalign{\smallskip}
Date & Telescope & Instrument & Duration & Spectral range  \\
& & & seconds & \AA \\[3pt]
\hline\noalign{\smallskip}
Dec 7, 1994 & BTA, Russia & Four-color photometer & 2400 & U + B + V + R \\ 
 & & with photomultipliers & & \\[3pt]
Dec 2, 1999 & WHT, Canary  & Avalanche photo-diode & 6600 & 4000-7500 \\
 & Islands & & & \\[3pt]
Jan 9, 2000 & BTA, Russia & Panoramic photometer & 7900 & U + B + V + R \\
& & with position-sensitive detector & & \\[3pt]
Nov 15, 2003 & BTA, Russia & Avalanche photo-diode & 1800 & 4000-7500 \\[3pt]
Jan 25, 2007 & BTA, Russia & Panoramic photometer & 10500 & B + R \\
Jan 26, 2007 & BTA, Russia & with position-sensitive detector & 6500 & B + R \\
\noalign{\smallskip}\hline
\end{tabular}
\end{table*}

We analyzed the sample of observational data obtained by our group over the
time span of 12 years on different telescopes. The details of
observations are summarized in Table \ref{table_observations}. The equipment used were
a four-color standard photometer with diaphragms based on
photomultipliers, a fast photometer with avalanche photo-diodes \citep{shearer}
and a panoramic photometer based on position-sensitive detector \citep{psd,mppp}. 
All devices are photon counters which record the photon time of
arrivals with accuracy better that at least 1$\mu$s, and the final
observational data are the lists of these times. In case of panoramic
photometer \citep{psd} only photons arrived in 3$"$ aperture around the pulsar
location has been considered.

The photon lists of all observational sets have been processed in the same way by using the same 
software to exclude the systematic differences due to data analysis
inconsistencies.
Photon arrival times of 1999, 2003 and 2007 years sets have been collected
with absolute time scale calibration by means of GPS receivers. So, they
have been corrected to the barycenter
of the Solar System using the adapted version of {\it axBary} code by 
Arnold Rots. The accuracy of this code has been tested with detailed examples
provided by \cite{lyne_bary_examples} and is found to be better than at least 2$\mu$s.

The barycentered photon lists then have been folded independently using both
Jodrell-Bank radio ephemerides \citep{jb_ephem} and
our own fast-folding based method of timing model fitting \citep{fast_fold}.

The declared accuracy of Jodrell-Bank ephemerides frequency and derivatives
provide the folding precision of at least 1 $\mu$s, and the base epoch -- of at
least 5 $\mu$s \citep{jb_ephem}, so we decided to fold the light curves with 5000
bin (6.6 $\mu$ s) resolution.

The observations of 1994 and 2000 have been performed without absolute time
scale calibration, using non-stabilized frequency generators, so these data
can't be converted to the barycenter correctly. So, we divided the whole data
set into several pieces short enough to be well fitted with a 3rd order timing
model (up to second frequency derivative), performed an independent timing
model fit for each, folded and combined it together, compensating the phase
shift between separate pieces. The accuracy of such procedure is proved to
provide similar time resolution, so we use the same number of bins in its
analysis.

The observations of 1994 and 2000 have been performed in standard
Johnson-Cousins U, B, V and R photometric bands, 2007 -- in B and R bands,
while 1999 and 2003 -- without filters, but with the same detector, described
by \cite{apd} (avalanche photodiode with broad spectral sensitivity in the
4000-7500\AA~ peaked at $\sim$ 7000\AA).

\section{Study of phase stability}
\label{sec_phase}

\begin{figure}
\centering
{\centering \resizebox*{1\columnwidth}{!}{\includegraphics{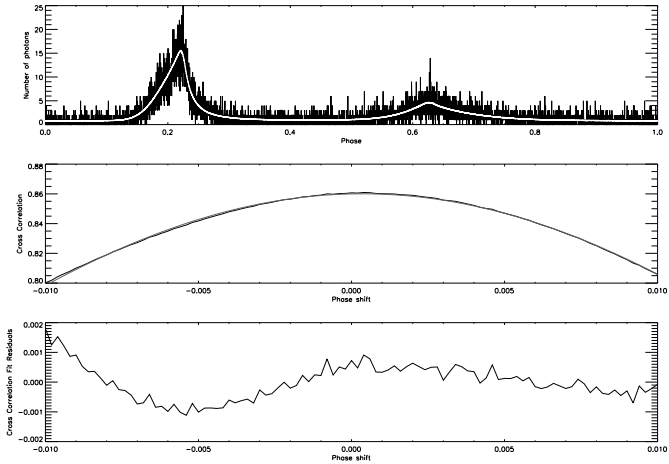}} \par}
\caption{Method of  estimation of phase shift of the sample folded light
  curve in respect to the template one. Upper panel -- sample light curve of 1999
  data set with 5000 bins resolution and 100 pulsar periods long
  (approximately 3.3 sec). Superimposed is the template profile, derived by
  folding the whole data set. Middle panel -- cross-correlation of the sample
  and the template light curves, with Gaussian fit superimposed. Lower panel --
  cross-correlation residuals after subtraction of Gaussian fit.  The real
  accuracy of phase shift estimation is much worse than the one of determining the
  cross-correlation peak position due to influence of original light curve
  errors and its systematic deviation from the Gaussian approximation. However,
  it may be shown that it does not lead to the statistical biasing of the
  estimate.}
\label{fig_shift_sim_single}
\end{figure}

\begin{figure}
\centering
{\centering \resizebox*{1\columnwidth}{!}{\includegraphics[angle=270]{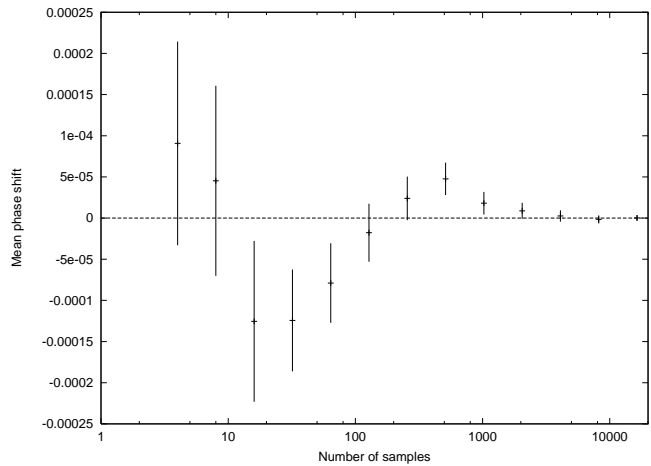}} \par}
\caption{Dependence of the estimation of phase shift mean value of simulated
  data on the set length. The simulated data have the same parameters (template
  profile, number of lightcurve bins and mean number of photons per sample) as
  the 1999 set. No significant biasing is seen, and the deviation from zero
  decreases with the increase of set length. It proves that the estimation of
  phase shifts is statistically unbiased and reliable.}
\label{fig_shift_sim_mean}
\end{figure}

\begin{figure}
\centering
{\centering \resizebox*{1\columnwidth}{!}{\includegraphics{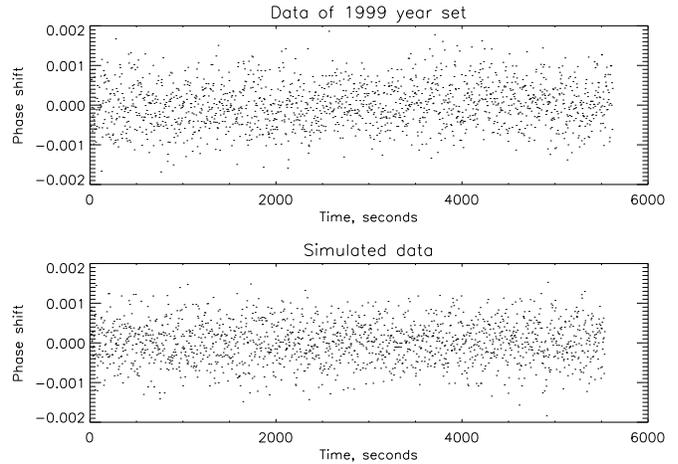}} \par}
\caption{Upper panel -- phase residuals of the Crab pulsar after applying the
  third-order timing model (up to second frequency derivative). It corresponds
  to the Gaussian noise with $5.5\cdot10^{-4}$ cycles rms. Lower panel --
  results of the similar analysis of the simulated data with the same mean
  parameters. The statistical properties are roughly the same as for the real
  data set.}
\label{fig_shift_1999}
\end{figure}

\begin{figure}
\centering
{\centering \resizebox*{1\columnwidth}{!}{\includegraphics[angle=270]{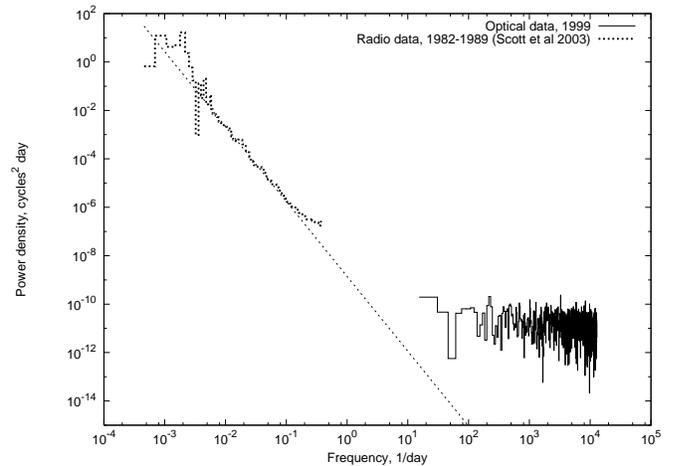}} \par}
\caption{Power density spectrum of the timing residuals of the 1999 data set,
  computed according to method described in \cite{scott}. Also, the spectrum of
  radio residuals from \cite{scott} is shown along with its power-law fit with
  a slope $\alpha=-3.09$.}
\label{fig_shift_1999_power}
\end{figure}

\begin{figure}
\centering
{\centering \resizebox*{1\columnwidth}{!}{\includegraphics[angle=0]{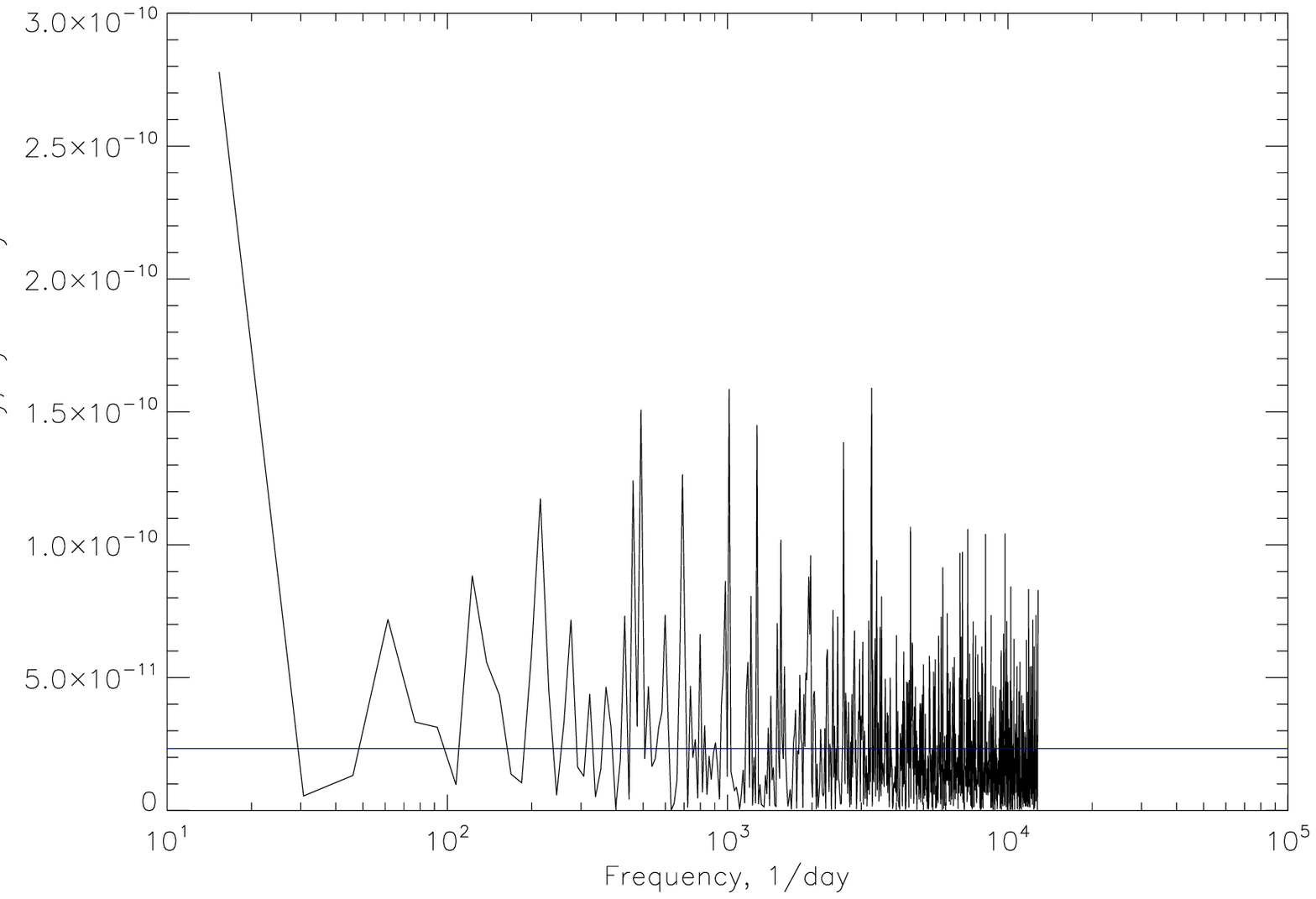}} \par}
\caption{Power density spectrum of the timing residuals of the 1999 data set,
  computed without any time domain window. The horizontal line shows the mean
  noise level. The deviation from it of the first bin value has the
  significance level $SL=2.4\cdot10^{-4}$ and suggests the presence of
  variations on time scale comparable with data set length.}
\label{fig_shift_1999_power_first}
\end{figure}

We performed the search for timing model residuals (``phase shifts'')
using two longest continuous data sets of 1999 and 2007 years.

\subsection{Computation of timing residuals}

The photon list of 1999  data set has been divided into segments of 100 pulsar
periods (with duration of approximately 3.3 s), which resulted in 1677 segments
with mean number of photons of 10546.5 each.  They then have been folded
separately using the same Jodrell-Bank radio ephemerides, and each fold have
been cross-correlated with the template, which has been built by folding the
entire data set. The estimation of the segment phase shift in respect to
template is then derived by approximating the peak of cross-correlation
function (in a phase window of 0.02 period width) with the Gaussian, and
analytically computing its maximum position. The steps of the procedure are
illustrated in Fig.~\ref{fig_shift_sim_single}. The resulting phase shifts are
plotted in upper panel of Fig.~\ref{fig_shift_1999}.

The formal accuracy of the maximum of Gaussian approximation of
cross-correlation function estimation is much better than the spreading of real
data values (basically, the error bars are hidden inside the dots in
Fig.\ref{fig_shift_1999}). This is partly due to neglecting of original light
curve errors while computing the cross-correlation. Also, the lower panel of
Figure~\ref{fig_shift_sim_single} demonstrates that the Gaussian estimation for
the cross-correlation function is not perfect, as it shows systematic
deviations from it. To ensure that these facts are not spoiling the results and
to test the statistical quality of the method used we performed the numerical
simulation by generating the set of sample Poisson-distributed light curves
based on the average profile of the 1999 data set with the same mean number of
photons per segment, and processing them in the same manner as the real
data. The computed phase shifts of simulated data are shown in the lower panel
of Fig.~\ref{fig_shift_1999}. The RMS of the simulated and observed data are
roughly the same ($\sigma \approx 5.5\cdot 10^{-4}$), which is consistent with
the statistical nature of phase shift values scatter. Also, we performed the
test whether the estimated phase shifts are unbiased and statistically reliable
by studying the behaviour of the mean value of simulated data phase shift and
its RMS in dependence of number of segments. The results of this simulation are
shown in Fig.~\ref{fig_shift_sim_mean}. It may be easily seen that the mean
value of the estimation along with its RMS, both converge to zero with the
increase of the number of segments, which proves the statistical reliability of
the method used. Also, it permits to increase the determination of systematic
phase shifts accuracy by averaging it over long phase segments.

\subsection{Fourier analysis of phase shifts}

In order to search for periodic components of the phase shift we performed the
Fourier analysis of the data set according to the method described in
\cite{scott}, i.e. computed the power-density spectrum $P$ using the
time-domain Hann window $w_i\propto\sin^2{\left(\pi i/N\right)}$ to suppress
the power leakage (which lowers the spectral resolution approximately by two).

The resulting power density spectrum is shown in
Fig.~\ref{fig_shift_1999_power} in comparison with results of radio data
analysis of \cite{scott}. The accuracy of our data is not sufficient to
reach the level of extrapolated power-law timing noise seen in the radio band.

However, it is possible to derive upper limits for the sinusoidal variable
components of timing noise on 3.3 s - 50 minutes time scale. Indeed, for the purely
white noise process the value $2P/\langle P \rangle$ is distributed as $\chi^2$
with 2 degrees of freedom \citep{leahy} and has mean and standard deviations of 2.
The probability $Q$ for this quantity to exceed some threshold value
$2P_0/\langle P \rangle$ by chance is
\begin{equation}
  Q\left(\chi_0^2=\frac{2P_0}{\langle P \rangle}\right) = \int\limits_{\chi_0^2}^{\infty}p(\chi^2)d\chi^2
\end{equation}
For the significance level $SL=0.01$ (which corresponds to the 99\% confidence probability)
the threshold value $P_0$, determined by solving
\begin{equation}
  Q\left(\chi_0^2=\frac{2P_0}{\langle P \rangle}\right) = c \mbox{\ ,}
\end{equation}
is $P_0=4.6\langle P \rangle$, where $\langle P \rangle$ is the mean ``noise''
level of power density spectrum. By combining it with the spectral amplitude
$P_A=\frac{A^2}{2\delta\nu}$ (where $\delta\nu = 1/T$ is the spectral
resolution) of sinusoidal signal, we have
\begin{equation}
  \frac{A^2}{2\Delta\nu} < 4.6 \langle P \rangle
\end{equation}
and for the upper limit for amplitude
\begin{equation}
  A < \sqrt{9.2 \langle P \rangle\Delta\nu}
\end{equation}

For our data, the mean ``noise'' level of power density is
$\langle P \rangle = 2.2\cdot10^{-11}$ cycles$^2\cdot$day and $\Delta\nu=15.3$ day$^{-1}$,
which leads to the limit for amplitude of periodic component of timing
residuals $A < 5.6\cdot10^{-5}$ cycles (1.8 $\mu$s) in this frequency range.

\subsection{Secular behaviour of phase shifts}

\begin{figure}[t]
{\centering \resizebox*{1\columnwidth}{!}{\includegraphics[angle=0]{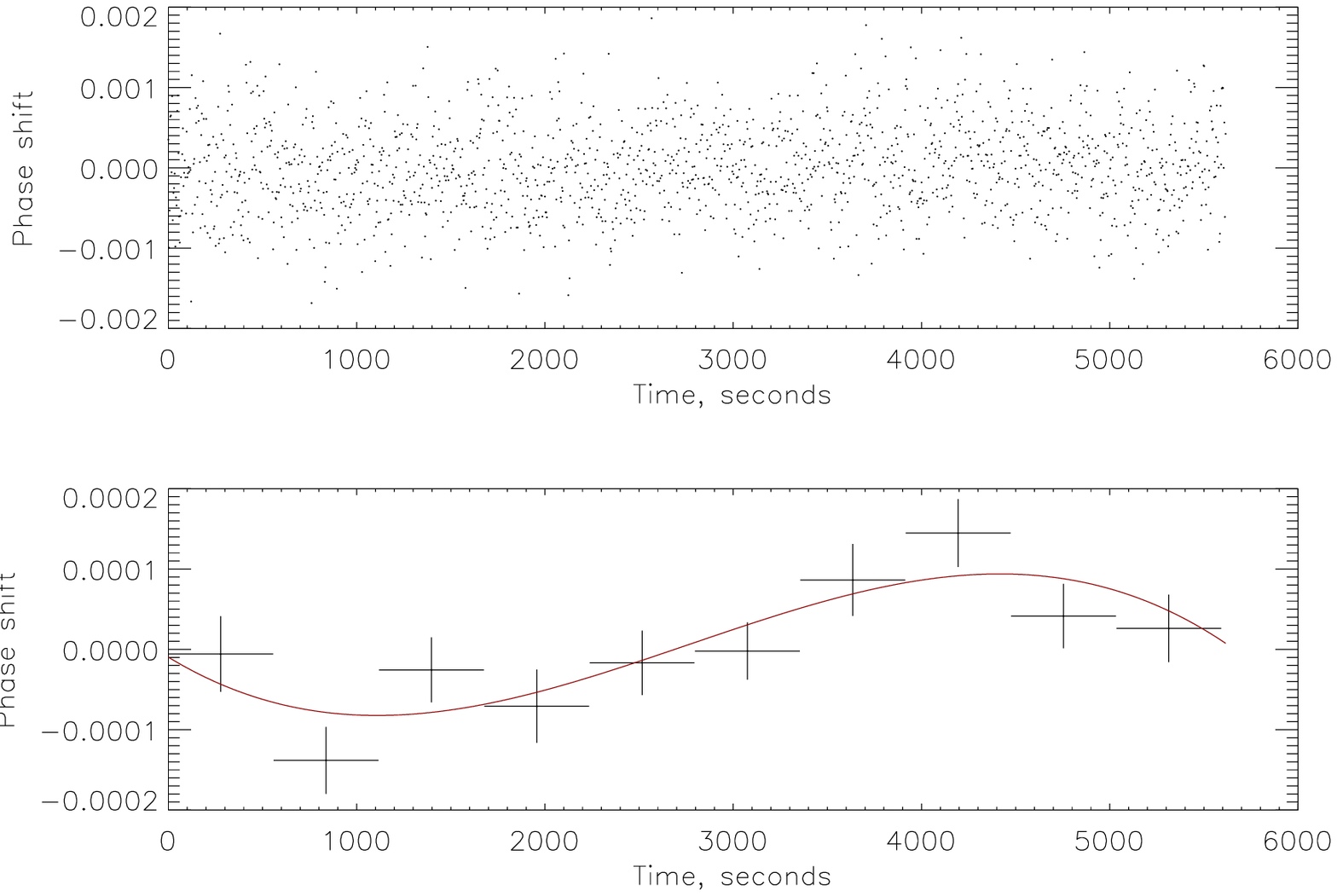}} \par}
\caption{Phase shifts of the 1999 year data set (upper panel, each point
  corresponds to 3.3 seconds of data) and its averaging in 10 time segments 562
  seconds long (lower panel). The vertical error bars correspond to the
  standard deviations of values in segments. Also, the approximation of phase
  shifts with 3-rd order polynomial. Characteristic time scale of variations is
  1.5-2 hours with amplitude $\sim$ $(1.5\pm0.5)\cdot 10^{-4}$ (4$\pm$1.5
  $\mu$s).}
\label{fig_crab_shift}
\end{figure}

\begin{figure}[t]
{\centering \resizebox*{1\columnwidth}{!}{\includegraphics[angle=0]{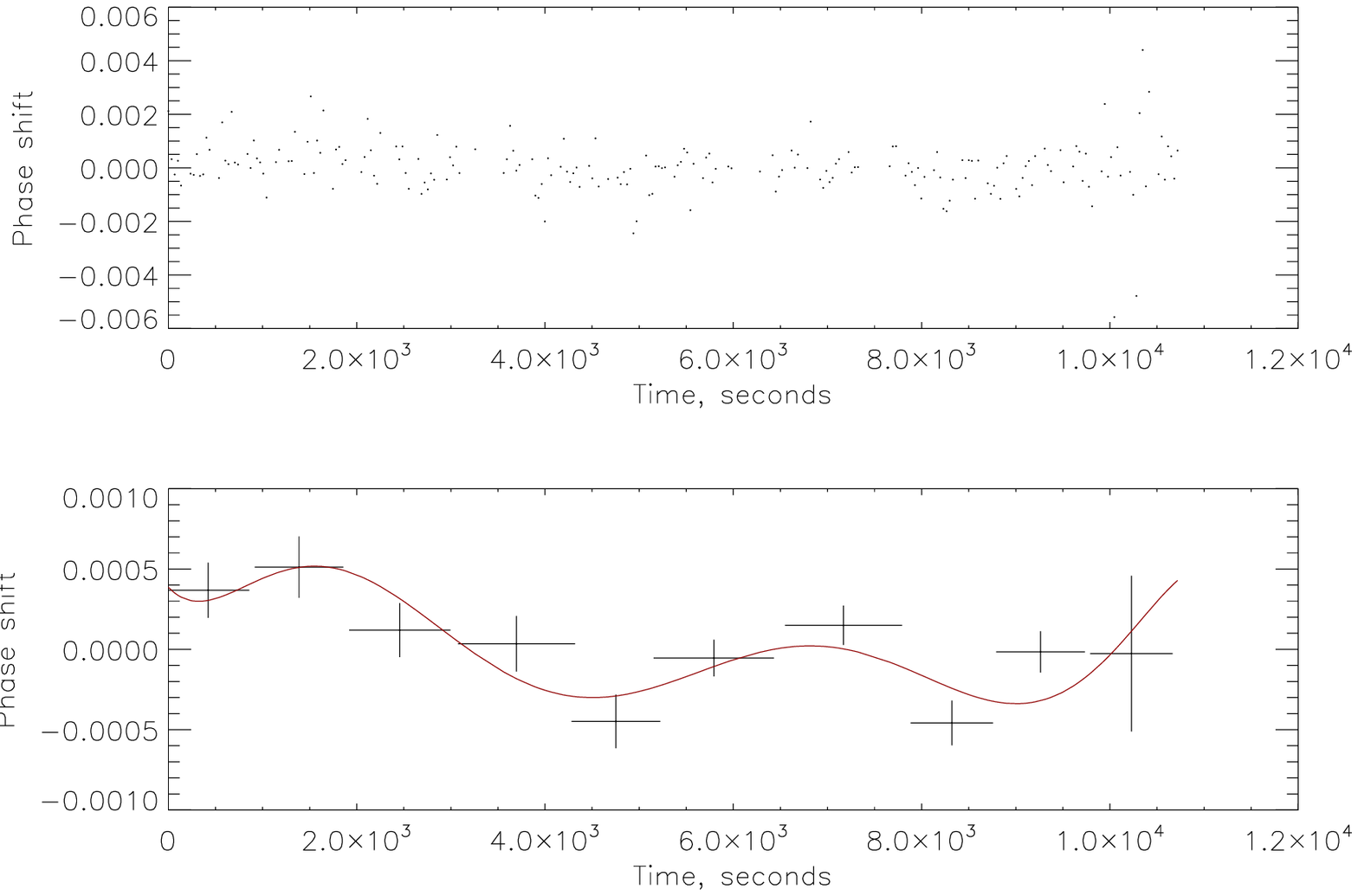}} \par}
\caption{Phase shifts of the Jan 25, 2007 data set (upper panel, each point
  corresponds to 33 seconds of data) and its averaging in 10 time segments 1050
  seconds long (lower panel). The vertical error bars correspond to the
  standard deviations of values in segments. Also, the approximation of phase
  shifts with 7-rd order polynomial is shown. Characteristic time scale of variations is
  1-2 hours with amplitude $\sim$ $(3\pm1)\cdot 10^{-4}$ (9$\pm$3
  $\mu$s).}
\label{fig_crab_shift_2007_1}
\end{figure}
\begin{figure}[t]
{\centering \resizebox*{1\columnwidth}{!}{\includegraphics[angle=0]{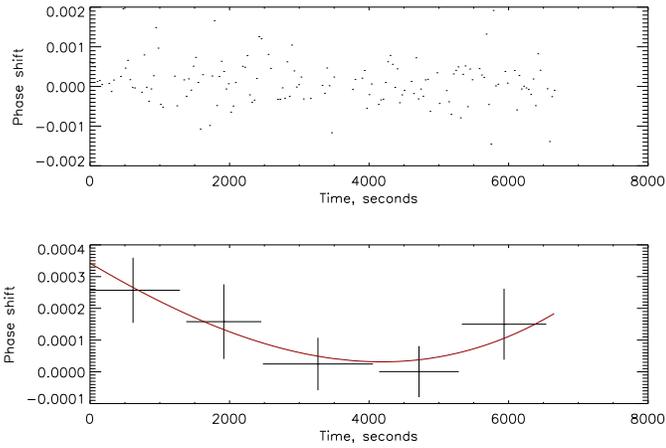}} \par}
\caption{Phase shifts of the Jan 26, 2007 data set (upper panel, each point
  corresponds to 33 seconds of data) and its averaging in 5 time segments 1300
  seconds long (lower panel). The vertical error bars correspond to the
  standard deviations of values in segments. Also, the approximation of phase
  shifts with 3-rd order polynomial is shown.}
\label{fig_crab_shift_2007_2}
\end{figure}

To test the phase stability of the Crab light curve on time scales comparable
with the total length of observations we first computed the power density
spectrum of phase shifts without any time domain window. The result, shown in
Figure~\ref{fig_shift_1999_power_first}, demonstrates significant deviation
(with significance level $SL=2.4\cdot10^{-4}$) of the first bin from noise mean
value, which suggests the presence of variations on time scale of the data set
length.

To check these variations in time domain we divided the data set into 10 equal
time segments 562 seconds long, and computed the mean and variance of phase
shifts in it. The results (shown in Figure~\ref{fig_crab_shift}) indeed show
the presence of significant variations on 1.5 hours time scale with
$(3\pm1)\cdot 10^{-4}$ (9$\pm$3 $\mu$s) amplitude.

We performed similar analysis of the data of Jan 25-26, 2007. The signal to
noise ratio in this set is smaller, so we computed the phase shifts using 1000
period long segments of light curve (roughly 33 seconds). The results (see
Figures~\ref{fig_crab_shift_2007_1} and \ref{fig_crab_shift_2007_2}) also show
variations on 1.5-2 hours time scale with similar amplitude.

\section{Pulse shape}

\begin{figure}[t]
{\centering \resizebox*{0.9\columnwidth}{!}{\includegraphics[angle=0]{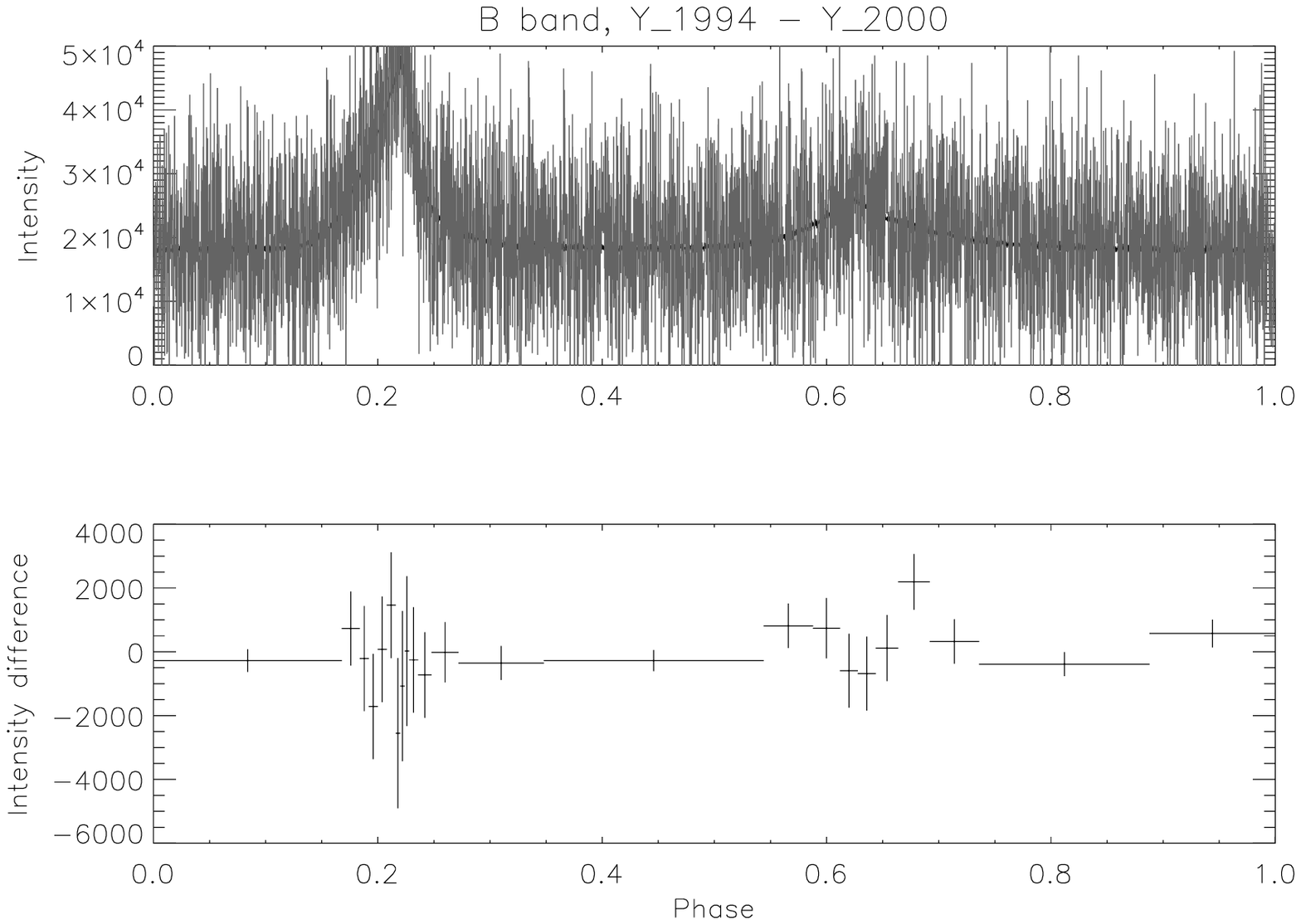}} \par}
{\centering \resizebox*{0.9\columnwidth}{!}{\includegraphics[angle=0]{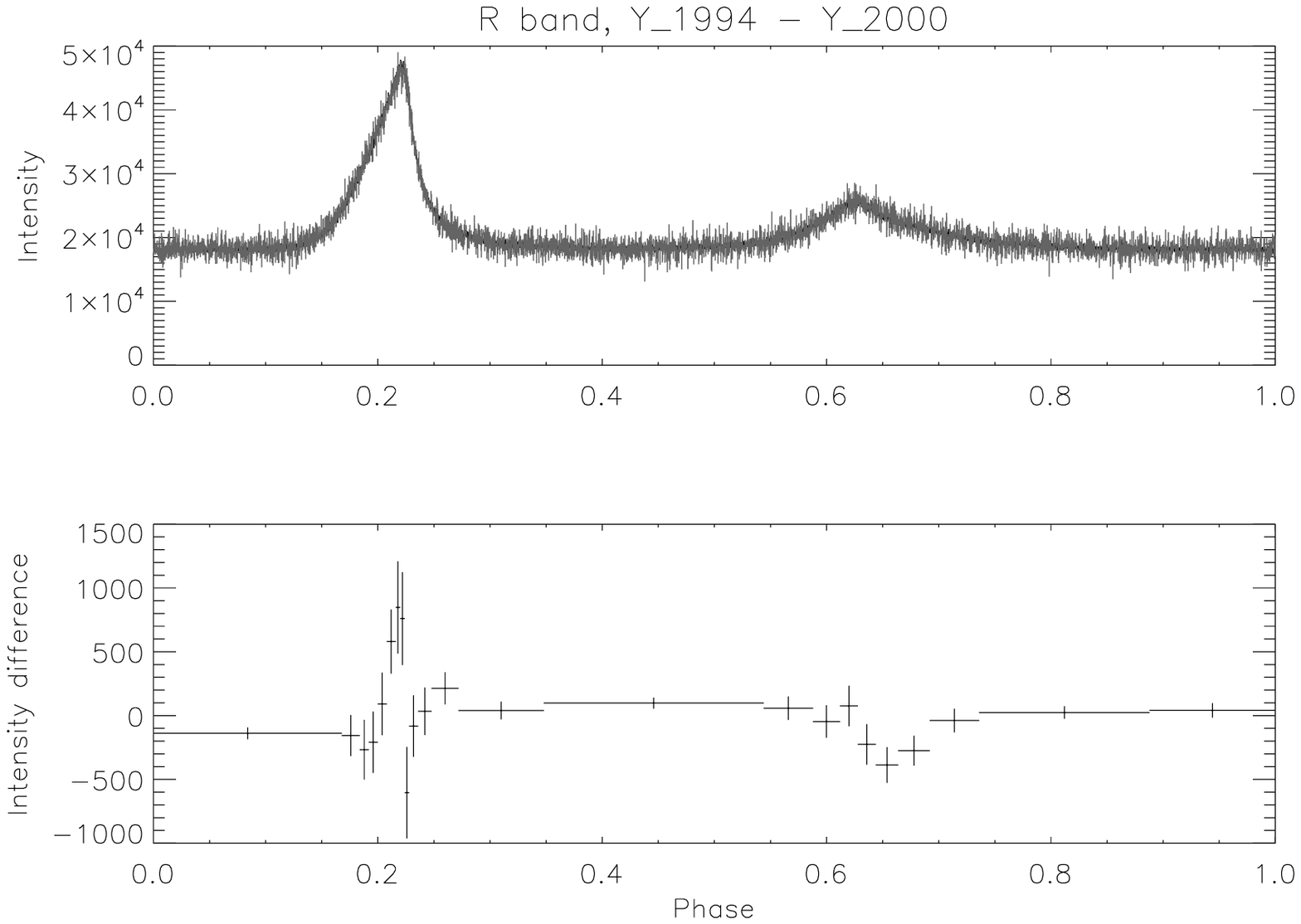}} \par}
\caption{Crab profile change in B and R filters between 1994 and
  2000. Intensity is in arbitrary units -- all light curves are normalized to
  the Nov 2003 data set one. R-band data demonstrates the combination of pulse
  phase shift ($\sim$ 30-50 $\mu$s) and small change of peaks right wings.
}
\label{fig_crab_change_1994_2000}
\end{figure}

\begin{figure}[t]
{\centering \resizebox*{0.9\columnwidth}{!}{\includegraphics[angle=0]{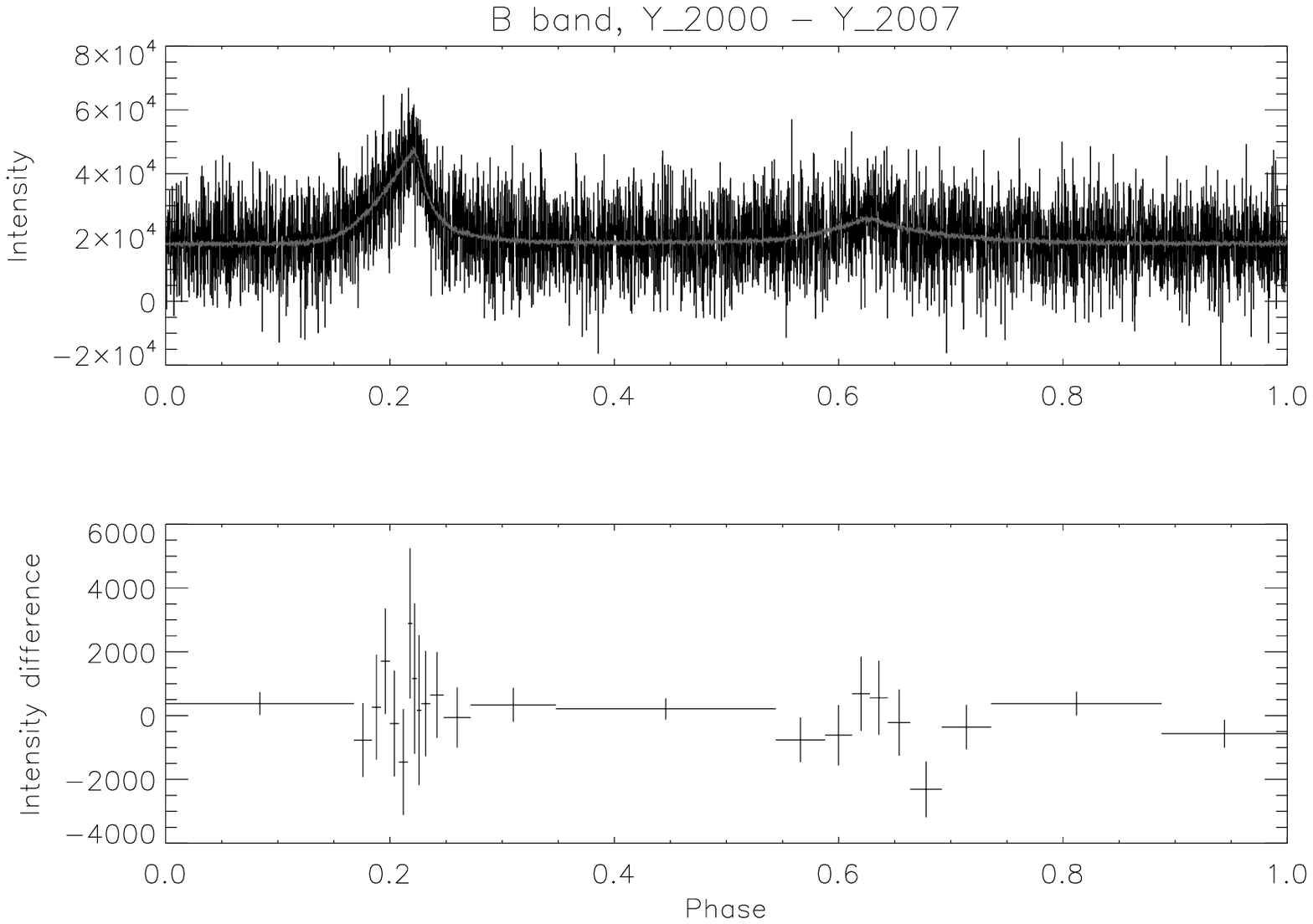}} \par}
{\centering \resizebox*{0.9\columnwidth}{!}{\includegraphics[angle=0]{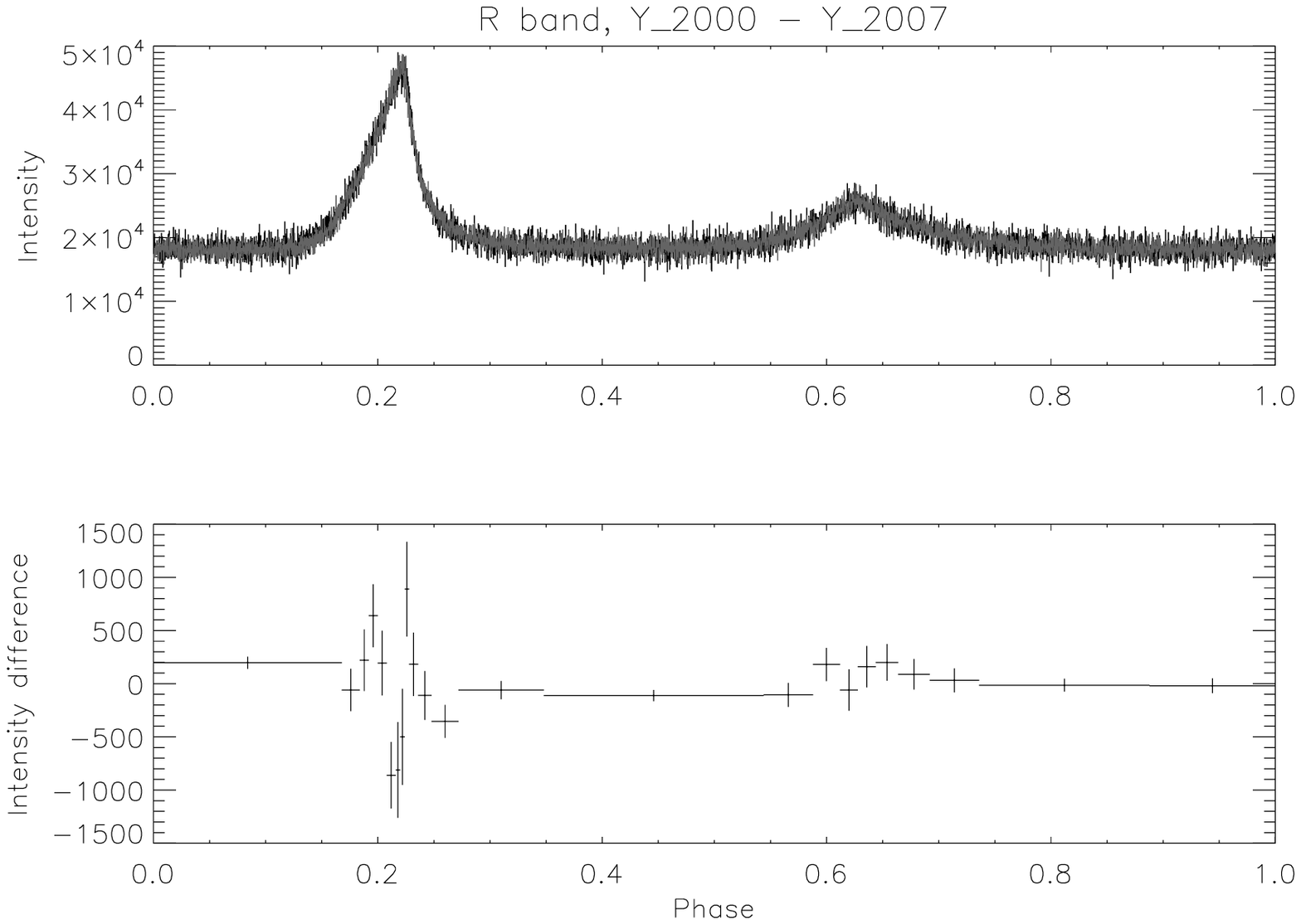}} \par}
\caption{Crab profile change in B and R filters between 2000 and 2007 data
  sets. Both bands demonstrate peaks (primarily -- first one) shape change
  without significant phase shift.
}
\label{fig_crab_change_2000_2007}
\end{figure}

\begin{figure}[t]
{\centering \resizebox*{0.9\columnwidth}{!}{\includegraphics[angle=0]{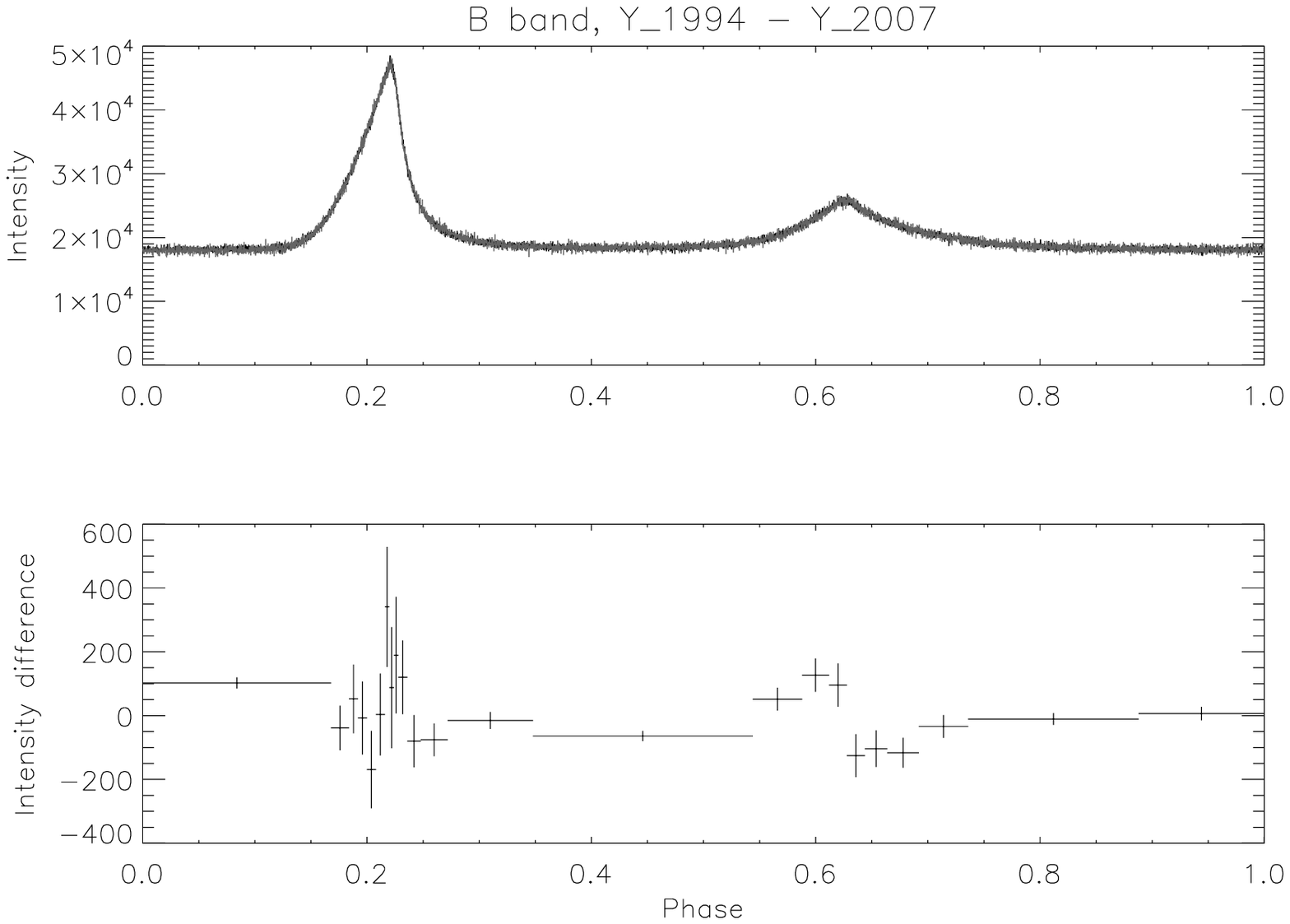}} \par}
{\centering \resizebox*{0.9\columnwidth}{!}{\includegraphics[angle=0]{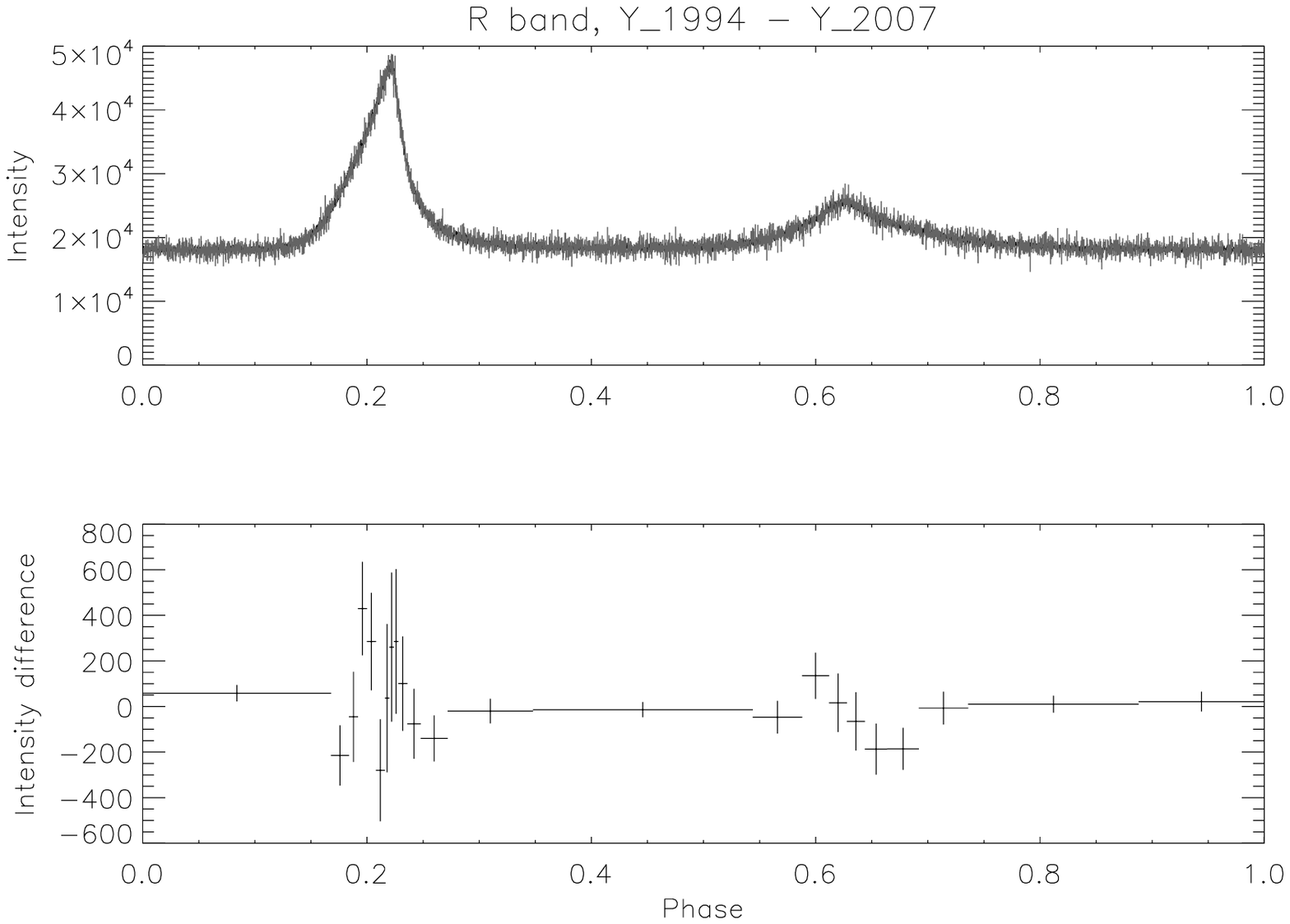}} \par}
\caption{Crab profile change in B and R filters between 1994 and 2007 data
  sets. Change of front wings of peaks is combined with the decrease of peak separation.
}
\label{fig_crab_change_1994_2007}
\end{figure}

\begin{figure}[t]
{\centering \resizebox*{0.9\columnwidth}{!}{\includegraphics[angle=0]{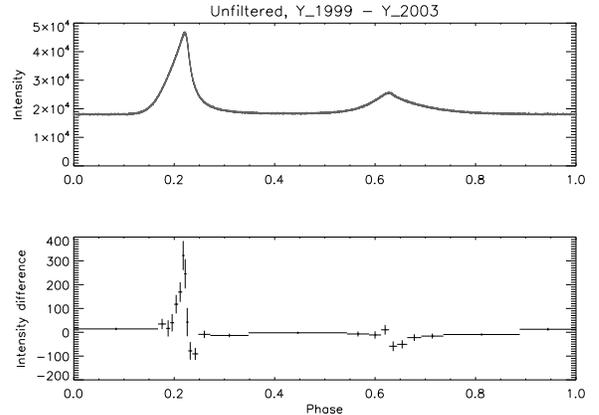}} \par}
\caption{Crab profile change between 1999 and 2003 data sets. The data are
  unfiltered, but acquired on the same photometer. Phase shift is combined with
  main peak right wing steepening.
}
\label{fig_crab_change_1999_2003}
\end{figure}

\begin{figure}
\centering
{\centering \resizebox*{1\columnwidth}{!}{\includegraphics{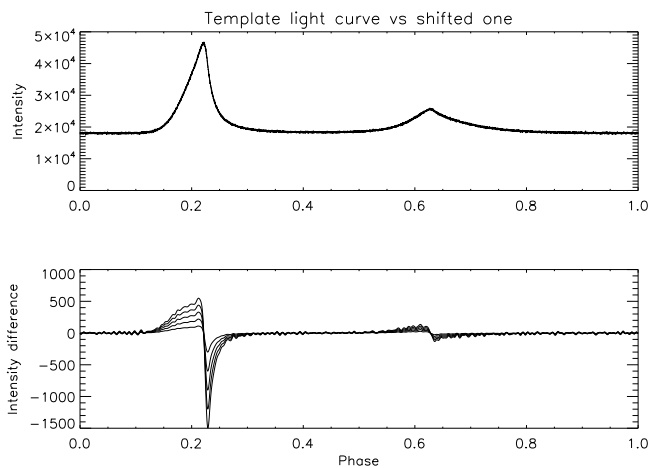}} \par}
\caption{Simulated residuals of the identical light curves with systematic
  shift of 1-5 bins (6.6 -- 33 $\mu$s). The shape of residuals is not like the
  ones discovered in comparison of different data sets.}
\label{fig_crab_change_shift}
\end{figure}

We performed the comparison of pulse profile shapes of data sets of 1994, 1999,
2000, 2003 and 2007 years in a way similar to the one used in
\citep{jones_1980}. As there are evidences that the Crab profile depends on
wavelength \cite{eikenberry_1996,golden_2000,beskin_2000,romani_2001}, we
compared the data acquired in the same wave bands, i.e. 1994 vs 2000 and 2007,
and 1999 vs 2003 data sets.

We compensated the phase shifts of 1999 and 2007 data described above by
approximating it with high-order polynomials and adding it to the timing
models. Also, to compensate possible phase shifts between light curves of
different data sets (due to, for example, systematic errors in radio
ephemerides base epochs) we determined the phase shift between them by the
method described in Section~\ref{sec_phase} and re-folded them with base epoch
shifted according to it. This procedure has been performed iteratively until
the phase shift became smaller than at least $10^{-4}$, i.e. less than half of a
bin size used.

Then we normalized each light curve $y_i$ to the same template profile
$y_{0,i}$ (we used the one of Nov 2003 data set as it has the largest number of
photons) by means of linear transformation $y_i'=ay_i+b$ with parameters $a$
and $b$ maximizing the likelihood
\begin{equation}
  L' = \sum_{i=0}^{N - 1} \left( y_i \ln{\lambda_i - \ln{(y_i!)} - \lambda_i}
  \right) \mbox{\ ,}
\end{equation}
of $y_i$ to be the instance of Poissonian distribution with
$\lambda_i = \frac{1}{a} y_{0, i} - \frac{b}{a}$.

Then we rebinned them in blocks with roughly equal number of photons
and plotted the difference between them. The results are presented in
Figures~\ref{fig_crab_change_1994_2000}-\ref{fig_crab_change_1999_2003}.

To check whether it may be due to uncompensated phase shift we simulated
shifted light curves, computed their differences and plotted them in
Figure~\ref{fig_crab_shift}. The simulated effect has different shape and
different ratio of positive and negative residuals, which argues for the
reality of the detected variation.

The effects seen in
Figs.\ref{fig_crab_change_1994_2000}-\ref{fig_crab_change_1999_2003} may be
interpreted as a some combination of systematic ``phase shift'', variation of
main peak shape and change of the distance between primary and secondary
peaks. Unfortunately, its exact nature can't be revealed by the methods used --
it is impossible to correctly define the ``phase shift'' of two profiles with
different shape, and the procedure of light curves phasing devours some part of
the shape change effect. However, it may only lower the significance of
detected residuals -- so the presence of the effect itself is undoubtful.

\section{Pulse fine structure}



\begin{figure}
\centering
{\centering \resizebox*{1\columnwidth}{!}{\includegraphics{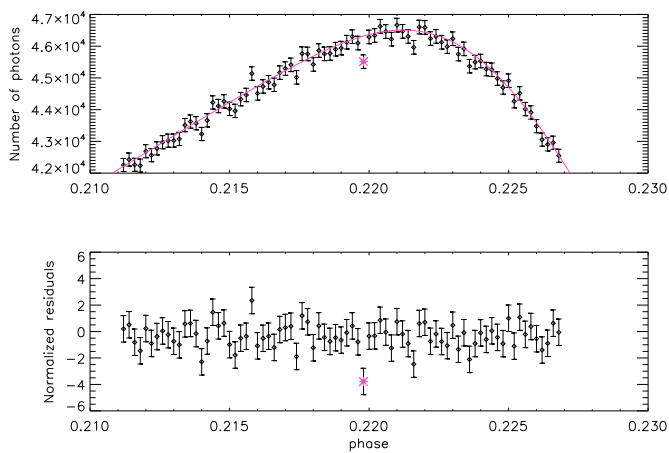}} \par}
\caption{Approximation of the peak of the Crab pulse profile of Nov 2003 data
  set by means of Fourier low-pass filter with characteristic time scale of 20
  bins (132 $\mu$s). The diagonal cross marks the feature with $A=3.77\sigma$
  amplitude.  The probability of its appearance on the peak (above the
  half-amplitude of the main pulse, 227 bins of 5000) by chance is 1.9\%.}
\label{fig_peak}
\end{figure}

For the data set of Nov 2003, which has the largest number of photons
collected, we performed the search for the fine structure of the peak of main
pulse. We approximated its shape by means of low-pass Fourier filter with
characteristic frequency of 0.05 bin$^{-1}$, which effectively smoothes the
light curve with sinc-like window with $\sim$ 40 bins FWHM. The peak data, fit
and residuals are shown in Figure~\ref{fig_peak}.  We do not detect any
significant spike-like fine structure on the level of 0.5\% (1 $\sigma$) with
6.6$\mu$s time resolution.

However, there is a single ``absorption-like'' feature with an amplitude of
$3.77\sigma$ near the maximum. The probability for such a feature to appear by
chance on the peak (on the segment with intensity above the half-amplitude
which contains 227 bins) is 1.9\%.



The light curves of other data sets, as well as their sum, however, do not show
any statistically significant deviations.

\section{Discussion}

Optical emission of Crab pulsar has been detected soon after its radio one
\citep{staelin_1968, cocke_1969}, and since then it has been observed a number
of times
\citep{Kristian, cocke_1974, peterson_1978, percival_1993, beskin_2000,
  golden_2000}).
Being the brightest ($\sim$16$^{\rm m}$) among 5 known optical pulsars, it
demonstrates relatively high stability of light curve shape against a
background predicted secular luminosity changes \citep{pacini, nasuti}. At the
same time, it exhibits the timing noise in pulse time of arrivals in radio,
optical and x-ray spectral bands on time scales from several days to tens of
years \citep{cordes_1980, boynton_1972, kuiper_2003, rots_2004}. Also, the
variation of primary and secondary peaks intensity ratio in gamma rays has
given the hint of 13-year periodicity \citep{nolan_1993, ulmer_1994}. Similar
behaviour has been observed in radio on months and years time scales
\citep{jones_1988, lyne_1988, scott}.  On the other hand, the search for time
of arrival residuals on short (seconds to hours) time scale has not been
practically performed. The one exception seems to be the result of
\citet{cadez_1996} and \citet{cadez_2001}, who detected the 60-seconds
periodicity of pulsar intensity.

We analyzed the data of several sets of optical observations with high temporal
resolution of the Crab pulsar performed by our group over the last 12 years.


No evidence for periodic short time scale variations of pulse time of arrivals
(like 60-sec free precession claimed by \citet{cadez_2001}) is detected on the
3.3 s -- 50 minutes time scale on Dec 2, 1999. The upper limit for their amplitude
is $A < 5.6\cdot10^{-5}$ cycles (1.8 $\mu$s) (significance level 0.01).  Note that 
no periodic features has also been detected in the Crab light curve on these
frequencies in previous work of \citet{golden_2000}.

Also, no signature of extended timing noise spectral features (like power-law
one observed on lower frequencies) is seen on this time scale.


However, the data of Dec 2, 1999 and Jan 25-26, 2007 sets both show significant
phase variations on 1.5 -- 2 hours time scale with $(2 - 5)\cdot10^{-4}$ cycles
(6 -- 16 $\mu$s) amplitude. This effect is most likely not truly
periodic. Moreover, it is difficult to explain it as a precession of a rotating
rigid body, as it requires too large difference of neutron star axes
($\Delta R/R\sim10^{-5}$) \citep{akgun_2006}.

Possible manifestations of noise processes, related to either superconducting
vortices inside the neutron star, or magnetospheric effects, on the short time
scales of minutes to hours has yet to be analyzed. These effects has been
usually involved in explanation of timing noise observed on time scales of days
to years \citep{cordes_1981, alpar_1986, cheng_1987a, cheng_1987b}. However,
the amplitude of the effect we discovered significantly exceeds the power-law
extrapolation of timing noise spectrum \citep{scott} (see
Fig.\ref{fig_shift_1999_power}).

Only one observed effect is currently known to occur on similar time scale --
the giant radio pulses \citep{lundgren_1995}, which have inverse power-law
intenstity distribution and randomly appear in all phases of light curve
occupied by ``normal'' radio emission except for precursor
\citep{jessner_2005}. Moreover, it has been recently suggested that all radio
emission except for the precursor consists of giant pulses only
\citep{popov_2006}. Their origin is most likely due to changes of either
coherence conditions or electron density in the magnetosphere. In the latter
case, it may influence the optical emission region too. The slight correlation
between giant radio pulses and increase of optical emission has been discovered
in \citep{shearer}. As giant pulses appear randomly in phase, they may lead to
changes of optical pulse shape, and so mimic the time of arrival variations.

Also, it may in principle be attributed to polar cap current-pattern drifting,
which may occur on a very broad range of time scales \citep{ruderman_2006}.

The non-detection of this effect in ongoing radio observations may be
attributed to its lower resolution (according to \citet{lyne_1993}, the
accuracy of pulse time of arrival determination in Jodrell Bank observations is 20
$\mu$s for 10 min integration time).


We discovered the variation of pulse shape between different sets of our
observations, i.e. on time scale of several years. It presents and has similar
properties in all studied spectral bands, and cannot be attributed to
well-known effect of shape dependence on wavelength
\citep{eikenberry_1996,golden_2000,beskin_2000,romani_2001}. Due
to limitations of data analysis methods used it is impossible now to specify
the exact nature of the variation -- it may only be empirically described as a
combination of systematic phase shift, main and secondary peaks shape change
and variation of peak separation. Also, it is not clear whether the
variation periodic, systematic or irregular. However, the effect is similar to
the one marginally detected in \citep{jones_1980} on time scale of 7 years.

There are several possible physical mechanisms able to produce such pulse shape
variations on time scale of years. First is the suspected precession of Crab on
$\sim$ 568 days \citep{scott}. Indeed, at least one other pulsar --
PSR~B1828-11 -- exhibits the precession accompanied by the changes of a radio
pulse profile on a similar time scale \citep{stairs}. For Crab, however, due to
difference in rotational frequencies, such precession period implies much
smaller wobble angle, and so -- smaller pulse profile variations. Also, polar
cap current-pattern drifting may mimic the precession and result in the same
phase shift and profile change effects on years time
scale.\citep{ruderman_2006}.

Another possibility is the incomplete post-glitch relaxation
\citep{demianski_1983, lyne_1993, lyne_2001}, as all our observations have been
separated by glitches of different power \citep{lyne_bary_examples}. In Crab,
it manifests itself as a persistent change of frequency derivative, and may be
attributed to small changes of the angle between the magnetic dipole and the
rotation axis \citep{link_1997, horvath_1997, link_1998}, which inevitably
leads to pulse profile variation. Also, pulse profile variations in hard energy
band are often observed in anomalous x-ray pulsars
\citep{kaspi_2003,morii_2004}, however, it is still not clear whether they
result directly from glitches.


We do not detect any spike-like fine structure of the main pulse maximum on the
level of 0.5\% (1 $\sigma$) with 6.6$\mu$s time resolution.

All the proposed explanations of discovered variations of pulse shape and time
of arrival are qualitative only and are in no sense complete. The observations
have to be continued, and the theoretical analysis still has to be
performed. We hope the study of such variations can help to elaborate the
theory of pulsar emission.

\section{Acknowledgements}
This work has been supported by the Russian Foundation for Basic
Research (grant No 04-02-17555), Russian Academy of Sciences (program
"Evolution of Stars and Galaxies"), INTAS (grant No 04-78-7366)
and by the Russian Science Support Foundation.


\end{document}